\documentclass[12pt]{article}

\usepackage[colorlinks,linkcolor=blue,citecolor=blue,urlcolor=blue,bookmarks,bookmarksnumbered]{hyperref}
\usepackage{amsmath,amssymb,amsfonts}
\usepackage[paper=letterpaper,margin=1.0in]{geometry}
\usepackage{graphicx,xcolor}

\usepackage{cite}

\newcommand{\be}{\begin{equation}}
\newcommand{\ee}{\end{equation}}
\newcommand{\bea}{\begin{eqnarray}}
\newcommand{\eea}{\end{eqnarray}}
\newcommand{\ba}{\begin{array}}
\newcommand{\ea}{\end{array}}
\newcommand{\ben}{\begin{enumerate}}
\newcommand{\een}{\end{enumerate}}
\newcommand{\bi}{\begin{itemize}}
\newcommand{\ei}{\end{itemize}}
\newcommand{\bc}{\begin{center}}
\newcommand{\ec}{\end{center}}
\newcommand{\bfig}{\begin{figure}}
\newcommand{\efig}{\end{figure}}
\newcommand{\bq}{\begin{quotation}}
\newcommand{\eq}{\end{quotation}}
\newcommand{\bt}{\begin{table}}
\newcommand{\et}{\end{table}}
\newcommand{\btab}{\begin{tabular}}
\newcommand{\etab}{\end{tabular}}
\newcommand{\bs}{\begin{slide}}
\newcommand{\es}{\end{slide}}

\newcommand{\IR}{\mathbb{R}}

\let\ba=\overline

\def\IR{\relax\leavevmode{\rm I\kern-.18em R}}
\def\ZZ{\relax\leavevmode
       \ifmmode\mathchoice
       {\hbox{\sf Z\kern-.4em Z}}
       {\hbox{\sf Z\kern-.4em Z}}
       {\lower.9pt\hbox{\scriptsize\sf Z\kern-.36em Z}}
       {\lower1.2pt\hbox{\tiny\sf Z\kern-.36em Z}}
       \else{\sf Z\kern-.4em Z}\fi}
\def\RR{\relax\leavevmode
       \ifmmode\mathchoice
       {\hbox{\sf R\kern-.4em R}}
       {\hbox{\sf R\kern-.4em R}}
       {\lower.9pt\hbox{\scriptsize\sf R\kern-.36em R}}
       {\lower1.2pt\hbox{\tiny\sf R\kern-.36em R}}
       \else{\sf R\kern-.4em R}\fi}

\def\resetby#1#2{\@addtoreset{#2}{#1}}
\def\seceq{\@addtoreset{equation}{section}
              \def\theequation{\thesection.\arabic{equation}}}

\def\Label#1{\label{#1}%
                \smash{\hbox to0pt{\raise1ex\hbox{\tiny[#1]}\hss}}}
\def\noLabels{\let\Label=\label}

\begin{document}

{\footnotesize
${}$
}

\bc

\vskip 1.0cm
{\Large \bf Snowmass 2022}\\
\vskip 0.2cm
{\Large \bf Infrared Properties of Quantum Gravity: \\
\vskip 0.1cm
UV/IR Mixing, Gravitizing the Quantum -}\\
\vskip 0.1cm
{\Large \bf Theory and Observation}\\
\vskip 1.0cm

\renewcommand{\thefootnote}{\fnsymbol{footnote}}

{\bf Per Berglund${}^{1}$,  \bf Laurent Freidel${}^{2}$,  \bf Tristan Hubsch${}^{3}$,}\\
{\bf Jerzy Kowalski-Glikman${}^{4, 5}$, Robert G. Leigh${}^{6}$, David Mattingly${}^{1}$,} \\
{ \bf Djordje Minic${}^{7}$, } \\

\vskip 0.5cm

{\it
${}^1$Department of Physics and Astronomy, University of New Hampshire, Durham, NH 03824, U.S.A. \\
${}^2$ Perimeter Institute for Theoretical Physics, 31 Caroline St. N., Waterloo ON, Canada  \\
${}^3$Department of Physics and Astronomy, Howard University, Washington, D.C.  20059, U.S.A. \\
${}^4$ University of Wroclaw, Faculty of Physics and Astronomy,
Pl. M. Borna 9, 50-204 Wroclaw, Poland \\
${}^5$ National Centre for Nuclear Research, Pasteura 7, 02-093 Warsaw, Poland \\
${}^6$ Illinois Center for Advanced Studies of the Universe and Department of Physics,
University of Illinois, 1110 West Green St., Urbana IL 61801, U.S.A. \\
${}^7$Department  of Physics, Virginia Tech, Blacksburg, VA 24061, U.S.A. \\
}

\ec

\vskip 1.0cm


\begin{abstract}
We discuss the possible appearance of several rather exotic phenomena in quantum gravity, including UV/IR mixing, novel modifications of infrared phenomenology that extend 
effective field theory approaches, and the relaxation of the usual notions of locality.
We discuss the relevance of such concepts in quantum gravity 
for quantum information science, cosmology 
and general quantum gravity phenomenology.
\end{abstract}


{\bf Contact person: Per Berglund, Per.Berglund@unh.edu}

\vspace{0.5cm}

{\bf  Theory Frontier Topical Groups}:\\
 {\bf  TF01}: String Theory, quantum gravity, black holes, and \\
 {\bf  TF10}: Quantum Information Science

\renewcommand{\thefootnote}{\arabic{footnote}}

\newpage

{\bf Introduction and Summary}:

A primary tool in high energy physics and other disciplines is effective field theory. As one goes to high energies, one encounters new poles in the scattering matrix, while low energy physics is governed by ``integrating out'' high energy modes. However, it is by no means obvious that key predictions in a quantum theory of gravity can be captured by effective field theory arguments. A simple example of why there may be a problem is provided by a thought experiment: in a gravitational theory, as we go to very high energies, we do not expect simply new poles in an S-matrix, but rather black hole formation. More generally, we may expect UV/IR mixing phenomena, with, in particular, new phenomena arising in the infrared as well as cosmological implications  \cite{dvali}.

Examples of UV/IR mixing in field theories are well-known, particularly in non-commuta\-tive field theories \cite{ncqft}, while its relevance in quantum gravity is currently an open question. A concrete framework for addressing this is to work within duality-symmetric formulations of string theory, which indeed possess some non-commutative features. For brevity we will refer to this approach as {\em metastring} theory \cite{Freidel:2015uug, Freidel:2016pls, Freidel:2013zga, Freidel:2014qna, Freidel:2015pka, Freidel:2017xsi,  Freidel:2018apz,  Freidel:2019jor, Minic:2020oho,  Freidel:2017wst, Freidel:2017nhg, Freidel:2021wpl}, and there is a useful phenomenological approach, called the metaparticle limit  \cite{Freidel:2018apz}, that captures some of the non-commutative features, leading to novel and interesting predictions. This theory can be understood as defining a precise notion of the {\em gravitization} of quantum geometry, in the sense that quantum mechanical structures that are traditionally fixed become dynamical.
Related cosmological studies making use of string dualities have also appeared in recent years, \cite{Brandenberger}, \cite{Swampland}
(see also,  \cite{Cohen:1998zx}).

What we are interested in here then are novel predictions at low energies which are unlike the usual signatures of quantum gravity theories (which typically strengthen with increasing energy). For example,  the metaparticle limit yields dispersion relations with distinctive infrared signatures \cite{Freidel:2021wpl}.
This novel and qualitatively different phenomenology is  relatively unexplored and will stimulate new observational approaches in quantum gravity phenomenology and should be further developed and pursued.  Additionally, quantum gravity models often give similar possible phenomenological effects despite drastically different theoretical underpinnings --- a classic and well-known example is the possible violations of Poincar\'e invariance, which has arisen in everything from causal sets~\cite{Dowker:2003hb}, to stringy spacetime foam backgrounds~\cite{Ellis:1999uh}, to loop quantum gravity~\cite{Gambini:1998it}, to Horava-Lifshitz gravity \cite{Horava:2009uw}.  Thus beyond the experimental motivation, one should understand whether these qualitatively new low energy effects can be naturally incorporated in other theories of quantum gravity, or whether they are unique to this particular (metastring/metaparticle) approach. This should be of interest both to theorists and phenomenologists.

Consistent with other non-commutative approaches, the metastring  mixes ultraviolet and infrared degrees of freedom, leading to infrared effects that cannot be captured by traditional Wilsonian effective field theory.  This yields another motivation for studying this theory, in that it provides a UV complete theory with UV/IR mixing and novel infrared phenomenology that evades standard  constraints that can be placed on the non-commutativity scale.~\cite{Szabo:2009tn}
This approach has novel implications for quantum information as well.  Deep connections between gravity and quantum information have recently been discovered, primarily in dual constructions with boundary quantum information constructs corresponding to bulk gravitational geometry (cf.~\cite{Ryu:2006bv,Almheiri:2014lwa, Brown:2015bva}).  However, the boundary quantum mechanics is ``un-disturbed'' in that it follows the usual rules for relativistic quantum theories. In metastring theory the presence of relative (or observer dependent) locality \cite{AmelinoCamelia:2011bm, AmelinoCamelia:2011pe}
implies that fundamental notions of quantum information theory, such as entanglement, that rely on factorization of the full quantum gravity Hilbert space into local subregions, can become observer dependent in a novel way that is distinct from standard QFT based localization prescriptions.  Additionally, one expects that fundamental quantum mechanical constructs, in particular the Born rule, will receive small corrections due to the gravitization of quantum geometry
\cite{Minic:2003en, Freidel:2014qna, Freidel:2016pls}, which could be explored through multi-path interference \cite{Sorkin:1994dt}.
While a small quantum gravity correction to the Born rule may appear innocuous and perhaps not even directly observable, due to the interlocking logical structure of quantum mechanics such a change can have profound cascading effects (e.g., failure of the principle of purification~\cite{Galley2018anymodificationof}) and therefore the theoretical ramifications should be fully explored.

{\bf Theoretical foundations:}
{One of the common features of various approaches to quantum gravity is the concept of
minimal length  \cite{Maggiore:1993rv, Kempf:1994su, Garay:1994en, Hossenfelder:2012jw, Chang:2011jj, Chang:2016sae}. Naively,  the presence of length or mass scales contradicts relativity (different boosted observers would see different minimal lengths/times, contradicting the very concept of ``minimal length/time'') \cite{Amelino-Camelia:2000stu,Amelino-Camelia:2000cpa,Rovelli:2002vp}.
It was discovered, however, that such theories could be made compatible with the relativity principle by deforming the Poincar\'e algebra; the price was that the concept of absolute locality (shared by all observers) does not survive in this context.
This observation was soon after reformulated as the principle of relative locality \cite{AmelinoCamelia:2011bm, AmelinoCamelia:2011pe}. The basic idea is that observations of spacetime structure can depend on properties of the probe, such as momentum, and the notion of locality becomes observer dependent. Simple models were constructed that possess  nontrivial geometry in momentum space, and more generally phase space. This principle relaxes a structure, that of locality, previously considered absolutely rigid (and fundamental to our usual notion of how the Lorentz group acts), similar to how relativistic physics relaxed the absolute time structure of Newtonian physics.}

Clearly  a fundamental property such as locality can only be modified very carefully. A natural question is whether the underlying conceptual idea of relative locality can be embedded in an established theory of quantum gravity.  The answer is yes, in that one can realize relative locality in string theory~\cite{Freidel:2015uug, Freidel:2016pls, Freidel:2013zga, Freidel:2014qna, Freidel:2015pka,  Freidel:2017xsi,  Freidel:2018apz,   Freidel:2019jor, Minic:2020oho,  Freidel:2017wst, Freidel:2017nhg, Freidel:2021wpl}.
This approach follows the line of reasoning well known from Einstein's special theory of relativity, albeit
in the quantum context.
In order to reconcile a minimal length with relativity one is led to covariant relative locality
\cite{AmelinoCamelia:2011bm, AmelinoCamelia:2011pe}, which in this setting means ``observer dependent spacetimes.''
Such observer dependent spacetimes
appear as ``slices'' in a new and larger entity  --- quantum spacetime, in analogy with observer dependent space and time
in special relativity which appear as ``slices'' in the larger, more general spacetime.
The new concept of quantum (or modular) spacetime \cite{Freidel:2015uug} is implied by a novel formulation of quantum physics
\cite{Freidel:2016pls} in terms of
unitary variables which explicitly contain fundamental length and time (such observables are directly related to
``modular variables'' \cite{AharonovRohrlich} (see also, \cite{Hooft:2014kka})).
The concept of modular spacetime also illuminates the problem of general quantum measurements.

It  turns out that non-locality (as defined by covariant relative locality) together with causality
essentially determines the known structure of quantum theory \cite{Freidel:2016pls}.
The geometry of such a causal non-local quantum spacetime is
called Born geometry \cite{Freidel:2013zga, Freidel:2014qna}, which unifies the symplectic, orthogonal and metric geometries.
Then, in analogy with Einstein's  theory of general relativity,
a dynamical (modular) quantum spacetime appears as the natural
arena for quantum gravity, gravitizing the quantum \cite{Freidel:2014qna}. This structure is precisely
realized in metastring theory, a T-duality covariant and intrinsically non-commutative formulation of string theory  \cite{Freidel:2015pka, Freidel:2017xsi}), and its
unitary and causal metaparticle limit \cite{Freidel:2018apz} . In this  setting quantum gravitational effects
appear both at short (UV) and  long distances (IR) \cite{Freidel:2019jor, Minic:2020oho}.

{\bf Observations:}
The observable hallmarks of this approach to quantum gravity are as follows:
the metaparticles are endowed with a non-trivial propagator and an IR sensitive dispersion relation, that
could be investigated in various experimental settings.
The metaparticles can be viewed as quanta of quantum fields in the modular representation.
Such fields are inherently bi-local and non-commutative, and each observed
quantum field has a ``dual'', which could be associated with dark matter degrees of
freedom \cite{Minic:2020oho}. Similarly, in the gravitational setting one gets, to leading order in the
minimal length, a dual geometry, which in the observed spacetime leads to ``dark energy'', modeled
as a positive cosmological constant \cite{Berglund:2019ctg, Berglund:2019yjq}.
(Dark matter, dark energy and visible matter are naturally correlated in this description
\cite{Ho:2010ca, Edmonds:2017zhg}.)
Also, the non-commutative nature of quantum spacetime provides a new interpretation of
axion-like backgrounds  \cite{Freidel:2017wst, Freidel:2017nhg},
as well as sheds light on the so-called generalized uncertainty principle
which is closely related to the idea of minimal length/time.

The most specific prediction of this approach is that in the deformed dispersion relation
(obtained in a particular gauge) $E_p^2 + \frac{\mu^2}{E_p^2} = {\vec{p}}^2 +m^2$,
the bound on the infrared scale parameter $\mu$ can be obtained to be (in the case of neutrinos)
of the order of $10^{-2} eV^2$ \cite{Freidel:2021wpl}. This is exciting, because it opens up a new window on
deformed dispersion relations which is found in the infrared (IR) domain, and it is within the reach of
current experiments. Another fascinating observation about this bound is that the characteristic energy
scale is close to the observed dark energy scale (roughly $10^{-3} eV$).
For comparison, note that tests of quantum gravity (and associated modified dispersion relations) from observations of gamma-ray bursts were suggested in \cite{AmelinoCamelia:1997gz}. Those modified dispersion relations are deformed in the UV,
 and since all bounds are essentially Planckian they are not easy to observe (see~\cite{AmelinoCamelia:1997gz} for the up to date review). In our case, crucially, we have a IR modification of the dispersion relation, which makes it possible for our proposal to become a new and interesting approach to quantum gravity phenomenology.
Ultrahigh-energy cosmic rays have been also suggested as probes of
possible violations of Lorentz invariance \cite{Liberati:2013xla}.
Once again, we emphasize that an effective violation of Lorentz invariance \cite{Mattingly:2005re} is encountered in the IR limit of
metaparticle physics \cite{Freidel:2018apz}, and it should be explored in this context as well.
Finally, the same line of investigation suggests interesting IR corrections to the traditional Yukawa-like scalar potentials.
%
All these avenues suggest a vibrant and diverse arena of physics effects associated with a new view on quantum gravity  that await to be explored in many
upcoming experimental searches \cite{Addazi:2021xuf}. These include gravitizing the quantum involving 
UV/IR mixing, covariant relative locality, Born
geometry, metastring theory and metaparticle physics, and thereby generalizing the standard notion of effective field theory.


\end{document}